\documentstyle[12pt]{article}
\begin{document}
\def\doublespaced{\baselineskip=\normalbaselineskip\multiply
    \baselineskip by 150\divide\baselineskip by 100}
\doublespaced
\pagenumbering{arabic}
\pagestyle{plain}
%
\begin{titlepage}
\begin{flushright}
{MSUHEP-50107\\
January, 1995 }
\end{flushright}
\vspace{0.4cm}
\begin{center}
\large
{\bf Effects of a heavy top quark on low energy data \\
     using the electroweak chiral Lagrangian}
\end{center}
\begin{center}
{\bf Ehab Malkawi and C.--P. Yuan}
\end{center}
\begin{center}
{Department of Physics and Astronomy \\
Michigan State University \\
East Lansing, MI 48824}
\end{center}
\vspace{0.4cm}
\raggedbottom
\setcounter{page}{1}
\relax

\begin{abstract}
\noindent
We study large top quark mass effects on low energy physics in
the chiral Lagrangian formulated electroweak theories.
We show that these radiative corrections can be easily obtained
from a set of Feynman diagrams involving only the scalar Goldstone
bosons and the fermions  when the contributions of
the order $g$ (the weak coupling) are ignored.
Using CERN LEP and SLAC Linear Collider (SLC) data we constrain on the
nonstandard couplings,
which might originate from the spontaneous symmetry-breaking sector,
of the top quark to the electroweak gauge bosons.
\end{abstract}

\vspace{2.0cm}

\end{titlepage}
\newpage
\section{Introduction}
\indent

The use of the effective Lagrangian does not necessarily stem from
our ignorance of the full dynamics.
In fact, as pointed out by H. Georgi in Ref.~\cite{Geor1}, the effective
field theory framework is not only simpler and more transparent, but
it actually provides a deeper insight into the relevant physics.
Commonly, the effective field theory approach
is applied for estimating the size of higher order
corrections \cite{Geor1}-\cite{wud}.
In this paper we consider an effective field theory approach
to electroweak radiative corrections and show
how it can conveniently relate various
radiative corrections  important for testing
the standard model (SM) in a rather elegant and clear way.
More importantly, this approach is shown to clearly identify
observables which are sensitive to the symmetry-breaking
sector of the electroweak theories.

The strong evidence \cite{D0}-\cite{alt}
for the presence of a heavy top quark motivates many studies on
various scenarios in which new physics shows up in the top quark
sector. For instance, in Ref.~\cite{ehab}, we studied
the general (nonstandard)
couplings of the top quark to the electroweak (EW) gauge bosons
in an effective chiral Lagrangian formulated electroweak theory with
the spontaneously broken symmetry
${\rm{SU(2)}}_{L}\times {\rm{U(1)}}_{Y}/{\rm{U(1)}}_{em}$.
The question regarding the origin of
such nonstandard interactions is of a great importance,
and was discussed to some extent in Ref.~\cite{ehab}.
Here, we shall not repeat that discussion but to recall our
conclusion that
the previously announced LEP data \cite{mele}
constrained somewhat the left-handed neutral current interaction.
On the other hand, the right-handed neutral current and the
charged currents were not usefully constrained by the data.

In this paper we will concentrate on two main points. The first point is
to study the leading
corrections of $O\left( m_t^2\ln \Lambda^2 \right)$
to low energy observables arising
from some nonstandard couplings of the top quark to
the EW gauge bosons, parameterized in the chiral Lagrangian.\footnote{
$\Lambda$ is the cutoff scale at which the effective
Lagrangian is valid.}
An earlier study was done in Ref.~\cite{ehab} by considering
a gauge invariant set of Feynman diagrams in which
massive gauge bosons appeared as external or
internal lines. The Goldstone bosons and top quark
only appeared as internal lines.
Because the leading radiative corrections
(in powers of $m_t$) are closely related to the
spontaneous symmetry-breaking (SSB) sector, we expect that
such corrections can be derived entirely from considering the interactions
between the top quark and the scalar bosons
(e.g., Goldstone bosons and possibly the Higgs boson).
In this paper we develop a different formalism to calculate these
leading corrections, using only the pure scalar boson and the
top-bottom fermionic
sectors in the chiral Lagrangian. We show how
to reproduce the results obtained in Ref.~\cite{ehab} from
a set of Feynman diagrams which only contain scalar boson
and fermion lines.

The second point is to update the constraints on the
nonstandard couplings of the top quark to the EW gauge bosons using the
new LEP \cite{lep94} and SLC data \cite{sld94}.
The rest of this paper is organized as follows. Sec. 2 is devoted to
study the large top quark mass contribution (in powers of $m_t$)
to low energy physics through the quantities $\rho$ and $\tau$ ~\cite{bar2}
in the chiral Lagrangian formulation.
In Sec. 3 we update the constraints on the
nonstandard couplings of the top quark to the EW gauge bosons.
(Previous constraints were given in
Ref.~\cite{ehab}.) Sec. 4 contains some of our conclusions.

\section{Large $m_t$ effects to low energy physics}
\indent

In this paper we are interested in the chiral Lagrangian formulated
electroweak theories in which the gauge symmetry
${\rm{SU(2)}}_{L}\times {\rm{U(1)}}_{Y}$ is nonlinearly realized.
{The relation between the linear and the nonlinear realization of
${\rm{SU(2)}}_{L}\times {\rm{U(1)}}_{Y}$\, corresponds to
some fixed nonlinear gauge transformation \cite{cole,chan}.}
The chiral Lagrangian approach has been used
in understanding the low energy strong interactions because it can
systematically describe the phenomena of spontaneous symmetry-breaking
\cite{wein}. Recently, the same technique has been widely
used in studying the electroweak sector
\cite{app,gold,fer,hol,pecc,how1,fer1}
to which this work has been directed.

A chiral Lagrangian can be constructed solely based upon the
broken symmetry of the theory, and it is not necessary to
specify the detailed dynamics of the actual breaking mechanism.
Hence, it is the most general effective Lagrangian
that can accommodate any underlying theory with that
pattern of symmetry-breaking at the low energy scale.
Furthermore, to study the low energy
behavior of such a theory using the chiral Lagrangian approach,
an expansion in powers of the external momentum
is usually performed \cite{geor2}.
In this approach, one generally considers
a Lie group $G$ which breaks down spontaneously
into a subgroup $H$. For every broken
 generator a Goldstone boson is introduced in the effective theory
\cite{cole}.
 In this paper we will concentrate on the electroweak group,
$G={\rm{SU(2)}}_L\times {\rm{U(1)}}_Y$ and $H={\rm{U(1)}}_{em}$.
There are three Goldstone bosons, $\phi^{a}$ ($a=1,2,3$),
generated by this breakdown, which are eventually eaten by $W^{\pm}$
and $Z$ gauge bosons and become their longitudinal degrees of freedom.

 The Goldstone bosons transform nonlinearly under $G$ but linearly
under the subgroup $H$. A convenient way to handle this is to introduce
the matrix field
\begin{equation}
\Sigma ={\rm{exp}}\left ( i\frac{\phi^{a}\tau^{a}}{v_{a}}\right )\, ,
\end{equation}
where $\tau^{a},\, a=1,2,3$ are the Pauli matrices normalized as
${\rm{Tr}}(\tau^a \tau^b)=2 \delta_{ab}$.
The matrix field $\Sigma$ transforms under $G$ as
\begin{equation}
\Sigma\rightarrow {\Sigma}^{\prime}={\rm {exp}}\left ( i
\frac{\alpha^{a}\tau^{a}}{2}\right )\,
\Sigma \,{\rm {exp}}(-i\frac{y\tau^3}{2})\, ,
\end{equation}
where $\alpha^{1,2,3}$ and $y$ are
the group parameters of $G$.
Because of the ${\rm{U(1)}}_{em}$ invariance,
$v_1=v_2=v$, but they are not necessarily equal to $v_3$.
In the SM, $v$ ($=246$\,GeV) is the vacuum expectation value of the Higgs
boson field, and characterizes the scale of the symmetry-breaking.
Also, $v_3=v$ arises from the approximate custodial symmetry
present in the SM.
It is this symmetry that is responsible for the
tree-level relation
\begin{equation}
\rho= \frac{M_W^2}{M_Z^2\,\cos^2 \theta_W}=1\,
\label{yeq3}
\end{equation}
in the SM, where $\theta_W$ is the electroweak mixing angle,
$M_W$ and $M_Z$ are the masses of $W^\pm$ and $Z$ boson, respectively.
In this paper we assume the full theory guarantees that
$v_1=v_2=v_3=v$.

It is convenient to define the composite fields
\begin{equation}
{{\cal W}_{\mu}^a}=-i{\rm{Tr}}(\tau^{a}\Sigma^{\dagger}
D_{\mu}\Sigma)\,
\end{equation}
and
\begin{equation}
{\cal B}_{\mu}=g^{\prime}B_{\mu}\,\, ,
\label{b}
\end{equation}
where
\begin{equation}
D_{\mu}\Sigma=\left (\partial_{\mu}-ig\frac{\tau^a}{2}W_{\mu}^a\right )
\Sigma\,\,\, .\end{equation}
In our notation $W_\mu^a$ and $B_\mu$ are the gauge bosons
associated with the ${\rm{SU(2)}}_L$ and ${\rm{U(1)}}_Y$ groups,
respectively. Also $g$ and $g'$ are the corresponding
gauge couplings.
The composite fields transform under $G$ as
\begin{equation}
 {{\cal W}_{\mu}^3}\rightarrow {{{\cal W}^{\prime}}_{\mu}^3}
       ={{\cal W}_{\mu}^3}-\partial_{\mu}y\, ,
\end{equation}
\begin{equation}
{{\cal W}_{\mu}^\pm}\rightarrow {{{\cal W}^{\prime}}_{\mu}^\pm}
  =e^{\pm iy}{{\cal W}_{\mu}^\pm}\, ,
\end{equation}
\begin{equation}
{\cal B}_{\mu} \rightarrow {\cal B}^{\prime}_{\mu} =
 {\cal B}_{\mu}+\partial_{\mu}y\,
\label{bb}\, ,
\end{equation}
where
\begin{equation}
{\cal W}_{\mu}^{\pm}={\frac{{\cal W}_{\mu}^{1}\mp i{\cal W}_{\mu}^{2}}
{\sqrt{2}}}\, .
\end{equation}

We also introduce the composite fields
${\cal Z}_{\mu}$ and ${\cal A}_{\mu}$ as
\begin{equation}
{\cal Z}_\mu={\cal W}^3_\mu +{\cal B}_\mu
\label{b1}\,\, ,
\end{equation}
\begin{equation}
s^2{\cal A}_\mu = s^2{\cal W}^3_\mu - c^2 {\cal B}_\mu\, ,
\label{b2}
\end{equation}
where $s^2\equiv\sin^2\theta_W$, and $c^2=1-s^2$.
In the unitary gauge ($\Sigma =1$)
\begin{equation}
{\cal W}_{\mu}^a=-gW_{\mu}^a \,\, ,
\end{equation}
\begin{equation}
{{\cal Z}}_{\mu} =-\frac{g}{c} Z_{\mu}\,\, ,
\end{equation}
\begin{equation}
{{\cal A}}_{\mu}=-\frac{e}{s^2}A_{\mu}\,\, ,
\end{equation}
where we have used the relations\, $e=g s=g' c$,
$W_\mu^3= c Z_\mu + s A_\mu$, and
$B_\mu= -s Z_\mu + c A_\mu$.
The transformations of ${\cal Z}_{\mu}$ and
${\cal A}_{\mu}$ under $G$ are
\begin{equation}
{\cal Z}_{\mu}\rightarrow {\cal Z}_{\mu}^{\prime}={\cal Z}_{\mu}\,\, ,
\end{equation}
\begin{equation}
{\cal A}_{\mu} \rightarrow {\cal A}_{\mu}^{\prime} =
{\cal A}_{\mu} -\frac{1}{s^2}\partial_{\mu}y \,\, .
\end{equation}
Hence, under $G$ the fields ${\cal W}_\mu^\pm$ and ${\cal Z}_\mu$
transform as vector fields, but ${\cal A}_\mu$ transforms as a gauge boson
field which plays the role of the photon field $A_\mu$.

Using the fields defined as above, one may construct
the ${\rm{SU(2)}}_L \times {\rm{U(1)}}_Y$ gauge invariant interaction
terms in the chiral Lagrangian
\begin{eqnarray}
{\cal L}^B =&-&\frac{1}{4g^2} {{\cal W}_{\mu \nu}^a}
{{\cal W}^{a}}^{\mu \nu}
 -\frac{1}{4{g^\prime}^2} {\cal B}_{\mu \nu}{\cal B}^{\mu \nu}\nonumber \\
&+&\frac{v^2}{4}{\cal W}^{+}_{\mu}{{\cal W}^{-}}^{\mu}+\frac{v^2}{8}
{\cal Z}_{\mu}{\cal Z}^{\mu}+{\dots }\,\, ,
\label{eq4}
\end{eqnarray}
where
\begin{equation}
{\cal W}^{a}_{\mu \nu}=\partial_{\mu}{\cal W}^{a}_{\nu}
-\partial_{\nu}{\cal W}^{a}_{\mu}+\epsilon^{abc}{\cal W}^{b}_{\mu}
{\cal W}^{c}_{\nu} \,\, ,
\end{equation}
\begin{equation}
{\cal B}_{\mu \nu}=\partial_{\mu}{\cal B}_{\nu}-\partial_{\nu}{\cal B}_{\mu}
\,\, ,\end{equation}
and where ${\dots}$ denotes other possible four-
or higher- dimensional operators \cite{app,fer}.

It is easy to show that\footnote{
Use ${\cal W}_{\mu}^a \tau^a = -2 i \Sigma^{\dagger} D_\mu \Sigma ~$,
and $[\tau^a,\tau^b]=2 i \epsilon^{abc} \tau^c $.}
\begin{equation}
{\cal W}_{\mu \nu}^a \tau^a=-g\Sigma^{\dagger}W^a_{\mu \nu}\tau^a
\Sigma\,\,
 \end{equation}
and
\begin{equation}
{{\cal W}_{\mu \nu}^a} {{\cal W}^{a}}^{\mu \nu}=g^2 W_{\mu \nu}^{a}
{{W^a}^{\mu \nu}}\,\, .
\label{eq05}
\end{equation}
This simply reflects the fact that the kinetic term is not related to
the Goldstone bosons sector, i.e., it does not originate from the
symmetry-breaking sector.

The mass terms in Eq.~(\ref{eq4}) can be expanded as
\begin{eqnarray}
\frac{v^2}{4}{\cal W}_{\mu}^{+}{{\cal W}^{-}}^{\mu}
+\frac{v^2}{8}{\cal Z}_{\mu}{{\cal Z}}^{\mu}
&=&{\partial}_{\mu}\phi^{+}\partial^{\mu}\phi^{-}
+\frac{1}{2}\partial_{\mu}\phi^{3}\partial^{\mu}\phi^{3} \nonumber \\
&&+\frac{g^2v^2}{4}W_{\mu}^{+}{W^{\mu}}^{-}
+\frac{g^2v^2}{8c^2}Z_{\mu}Z^{\mu}+{\dots}\,\, .
\end{eqnarray}
At the tree-level, the mass of $W^\pm$ boson is $M_W=gv/2$ and
the mass of $Z$ boson is $M_Z=gv/2c$.
The above identity implies that the radiative corrections to the
mass of the gauge bosons can be related to the wavefunction
renormalization of the Goldstone bosons, cf. Eq.~(\ref{wzmass}), and
therefore sensitive to the symmetry-breaking sector.

Fermions can be included in this context by assuming that each flavor
transforms under $G={\rm{SU(2)}}_L\times {\rm{U(1)}}_{Y}$ as \cite{pecc}
\begin{equation}
f\rightarrow {f}^{\prime}=e^{iyQ_f}f \label{eq1} \, ,
\end{equation}
where $Q_{f}$ is the electric charge of $f$.

 Out of the fermion fields $f_1$, $f_2$ (two different flavors),
and the Goldstone bosons matrix field $\Sigma$,
the usual linearly realized fields
$\Psi$ can be constructed. For example, the left-handed
fermions [${\rm{SU(2)}}_L$ doublet] are
\begin{equation}
\Psi_{L} = \Sigma F_{L} = \Sigma{f_1\choose {f_2}}_{L} \label{psi} \,
\end{equation}
with $Q_{f_1}-Q_{{f_2}}=1$.
One can easily show that $\Psi_{L}$\,transforms linearly under $G$ as
\begin{equation}
\Psi_{L}\rightarrow {\Psi}^{\prime}_{L}={\rm g} \Psi_{L}\, ,
\end{equation}
where ${\rm{g}}=
{{\rm {exp}}}(i\frac{\alpha^{a}\tau^{a}}{2})
{{\rm {exp}}}(i\frac{y}{2})\in G $.
Linearly realized right-handed fermions
$\Psi_{R}$  [${\rm{SU(2)}}_L$ singlet] simply coincide with $F_{R}$, i.e.,
\begin{equation}
\Psi_{R}= F_{R}={f_1\choose {f_2}}_{R}\, .\label{psr}
\end{equation}
It is then straightforward to construct
a chiral Lagrangian containing both the bosonic and the fermionic
fields defined as above .

Our goal is to study the large Yukawa corrections to
the low energy data from the chiral Lagrangian formulated
electroweak theories.
We shall separate the radiative corrections as an expansion in
both the Yukawa coupling $g_t$ and the weak coupling $g$.
($g_t=\sqrt{2} m_t /v$, where $m_t$ is the mass
of the top quark.)
With this separation we can then consider the case where
corrections of the order
$g$ are ignored compared to those of $g_t$.
This kind of study was done for the SM in Ref.~\cite{bar2},
where the gauge bosons were considered as
classical fields so that the full gauge invariance
of the SM Lagrangian was maintained,
and a set of Ward identities was derived to relate the Green's
functions of the Goldstone bosons and the gauge bosons.
Hence, large $g_t$ corrections can be easily obtained from
calculating Feynman diagrams involving only fermions (top and bottom
quarks)  and
scalar bosons (e.g., Goldstone bosons and possibly
the Higgs boson) but not gauge bosons.
The same conclusion can be drawn using the
chiral Lagrangian approach in a far more elegant and clear
way, as shown  below in this section.

Why is the chiral Lagrangian formulation useful in finding large
$g_t$ corrections beyond the tree-level?
In general  to perform a loop calculation, one needs to
fix a gauge and therefore explicitly
destroys the gauge invariance
$\left [ {\rm{SU(2)}}_L \times {\rm{U(1)}}_Y\right ]$  of the Lagrangian.
However, to find the large $g_t$ corrections
one does not need to include gauge bosons in loops \cite{bar2}. Thus,
there is no need to fix a gauge and the
full gauge invariance of the effective Lagrangian is maintained.
Because the chiral Lagrangian possesses the
${\rm{SU(2)}}_L \times {\rm{U(1)}}_Y$
invariance (nonlinearly) and the ${\rm{U(1)}}_{em}$ invariance (linearly)
at any given order of the perturbative expansions, and
all the loop corrections
can be reorganized using the composite fields ${\cal W}^{\pm}_{\mu},
{\cal Z}_{\mu}$, and ${\cal A}_{\mu}$ in a gauge invariant form,
therefore, it is the most convenient and elegant way to find
$g_t$ corrections beyond the tree-level.
This is obvious because the
leading radiative corrections (in powers of $m_t$)
are products of the SSB and therefore independent of the
weak gauge coupling $g$. We note that
in the expansion of the field
\begin{equation}
{\cal Z}_{\mu}=\frac{2}{v}\partial_{\mu}\phi^3 -\frac{g}{c}Z_{\mu}+...\, ,
\label{ex}
\end{equation}
there is always a factor $g$ associated with a weak gauge boson field.
Hence, loop  corrections independent of the gauge coupling
$g$ can be obtained by simply considering the scalar
and the fermionic sectors in the theory.
In the following we shall show how this is done.

\subsection{Effective Lagrangian}
\indent

To obtain the large contributions of the top quark
mass (in powers of $m_t$) to low energy data, we need only to
concentrate on the top-bottom fermionic sector ($f_1=t$ and $f_2=b$)
in addition to the bosonic sector.
The most general gauge invariant chiral Lagrangian can be written as
\begin{eqnarray}
{\cal L}_0&=&i\overline{t}\gamma^{\mu}\left ( \partial_{\mu}
 +i\frac{2s_0^2}{3}{\cal A}_{\mu}\right) t
+i\overline{b}\gamma^{\mu}\left (\partial_{\mu}-i\frac{s_0^2}{3}
{\cal A}_{\mu}\right ) b\nonumber \\
&&-\left (\frac{1}{2}-\frac{2s_0^2}{3}+\kappa_{L}^{\rm {NC}}\right)
\overline{t_{L}}\gamma^{\mu} t_{L}{{\cal Z}_{\mu}}
 -\left ( \frac{-2s_0^2}{3}+\kappa_{R}^{\rm {NC}}\right ) \overline{{t}_{R}}
\gamma^{\mu} t_{R}{{\cal Z}_{\mu}} \nonumber \\
&&-\left( \frac{-1}{2}+\frac{s_0^2}{3}\right)
\overline{b_{L}}\gamma^{\mu} b_{L}{{\cal Z}_{\mu}}
-\frac{s_0^2}{3}\overline{b_{R}}\gamma^{\mu} b_{R}
{{\cal Z}_{\mu}}\nonumber \\
&&-\frac{1}{\sqrt{2}}\left (1+\kappa_{L}^{\rm {CC}}\right ) \overline{{t}_{L}}
\gamma^{\mu} b_{L}
{{\cal W}_{\mu}^+}-\frac{1}{\sqrt{2}}\left
(1+{\kappa_{L}^{\rm {CC}}}^{\dagger}\right)
\overline{{b}_{L}}\gamma^{\mu}t_{L}{{\cal W}_{\mu}^-} \nonumber \\
&&-\frac{1}{\sqrt{2}}\kappa_{R}^{\rm {CC}}
\overline{{t}_{R}}\gamma^{\mu} b_{R}
{{\cal W}_{\mu}^+}-\frac{1}{\sqrt{2}}{\kappa_{R}^{\rm {CC}}}^{\dagger}
 \overline{{b}_{R}}\gamma^{\mu} t_{R}{{\cal W}_{\mu}^-} \nonumber \\
&&-m_t \overline{t} t +{\dots}
 \label{eq2} \,\, ,
\end{eqnarray}
where $\kappa_{L}^{\rm {NC}}$, $\kappa_{R}^{\rm {NC}}$,
$\kappa_{L}^{\rm {CC}}$, and $\kappa_{R}^{\rm {CC}}$
parameterize possible deviations from the SM predictions~\cite{ehab},
and ${\dots }$ indicates possible
Higgs boson interactions and other higher dimensional operators.
Here we have assumed that new physics from the SSB sector might
modify the interactions of the top quark to the EW gauge bosons.
On the other hand, the
bare $b$-$b$-$Z$ couplings are not modified in the limit of ignoring
the mass of the bottom quark \cite{ehab}.
The subscript $0$ denotes bare quantities and
all the fields in the  Lagrangian ${\cal L}_0$,
Eq.~(\ref{eq2}), are bare fields.

Needless to say, the composite fields are only used to
organize the radiative corrections in the chiral Lagrangian.
To actually calculate loop corrections one should expand these
operators in terms of the Goldstone boson and the gauge boson fields.
The gauge invariant result of loop calculations can be written
in a form similar to Eq.~(\ref{eq2}).
Denoting the fermionic part of this effective Lagrangian
as ${\cal L}_{eff}$, then
\begin{eqnarray}
{\cal L}_{eff} &=&i Z_b^{L} {{\overline{b_L}\gamma_{\mu}
\partial^{\mu} b_{L}}\,} +Z_1\frac{s_0^2}{3}{{\overline{b_L}
\gamma_{\mu} b_{L}{\cal A}^{\mu}}\,}
+\frac{1}{2}\left( Z_{v}^{L}-Z_2\frac{2s_0^2}{3}\right )
{{\overline{b_L}\gamma_{\mu} b_{L}{\cal Z}^{\mu}}\,}\nonumber \\
&&+iZ_b^{R}{{\overline{b_R}\gamma_{\mu}
\partial^{\mu} b_{R}}\,}+Z_3\frac{s_0^2}{3}{{\overline{b_R}
\gamma_{\mu} b_{R}{\cal A}^{\mu}}\,}
-Z_4\frac{s_0^2}{3}{{\overline{b_R}\gamma_{\mu} b_{R}{\cal Z}^{\mu}}\,}
+{\dots} \label{eq3} \, ,
\end{eqnarray}
in which  the coefficient functions $Z_1$, $Z_2$, $Z_{3}$, $Z_{4}$,
$Z_b^{L}$, $Z_b^{R}$, and  $Z_{v}^{L}$
contain all the loop corrections, and all the fields
in ${\cal L}_{eff}$ are bare fields.

In the case of ignoring the corrections of the order $g$, the
effective Lagrangian can be further separated into two
parts: one part has the explicit linear ${\rm{U(1)}}_Y$ symmetry
in the unitary gauge, and the other part contains all the radiative
corrections which do not vanish when taking the $g \rightarrow 0$
limit. Specifically, in this approximation, we can write
\begin{eqnarray}
{\cal L}_{eff} &=&i Z_b^{L} {{\overline{b_L}\gamma_{\mu}
\partial^{\mu} b_{L}}\,} -Z_1\frac{1}{3}{{\overline{b_L}
\gamma_{\mu} b_{L}{\cal B}^{\mu}}\,}
+\frac{1}{2} Z_{v}^{L} {{\overline{b_L}\gamma_{\mu} b_{L}
{\cal Z}^{\mu}}\,}\nonumber \\
&&+iZ_b^{R}{{\overline{b_R}\gamma_{\mu}
\partial^{\mu} b_{R}}\,}-Z_3\frac{1}{3}
{{\overline{b_R}\gamma_{\mu} b_{R}{\cal B}^{\mu}}\,} +{\dots}
 \label{eq3prime} \, ,
\end{eqnarray}
where
\begin{equation}
{\cal B}_{\mu}=s_0^2({\cal Z}_{\mu}-{\cal A}_{\mu})
\,\, ,
\end{equation}
derived from Eqs.~(\ref{b1}) and (\ref{b2}).
Note that as shown in Eqs.~(\ref{b}) and (\ref{bb}) the
field ${\cal B}_{\mu}$ is not composite and
transforms exactly like $B_\mu$.
Comparing Eq.~(\ref{eq3}) with (\ref{eq3prime}), we conclude that
the coefficient functions
 $Z_1$, $Z_2$, $Z_{3}$, and $Z_{4}$ must be related and
\begin{equation}
Z_2=Z_1\,\, ,
\end{equation}
\begin{equation}
Z_4=Z_3\,\, .
\end{equation}
All the radiative corrections to the vertex $b$-$b$-$\phi^3$
in powers of $m_t$ are summarized by
the coefficient function $Z_{v}^{L}$ because, from Eq.~(\ref{ex}),
\begin{equation}
\frac{1}{2}Z_{v}^{L}{{\overline{b_L}\gamma_{\mu} b_{L}
{\cal Z}^{\mu}}\,}=Z_{v}^{L}{\frac{1}{v}}
{\overline{b_L}}\gamma_{\mu}b_L \partial^{\mu}\phi^3 +{\dots }\,\,.
\end{equation}

Since the effective Lagrangian ${\cal L}_{eff}$
possesses an explicit ${\rm{U(1)}}_{em}$ symmetry and under $G$ the
field ${\cal A}_\mu$ transforms as a gauge boson field and
 ${\cal Z}_\mu$ as a neutral vector boson field,
therefore, based upon the Ward identities in QED
we conclude that in Eq.~(\ref{eq3})
\begin{equation}
Z_1 = Z_b^{L}\,\, ,
\end{equation}
and
\begin{equation}
Z_3=Z_b^{R}\,\, .
\end{equation}
Hence, the effective Lagrangian ${\cal L}_{eff}$ can be rewritten as
\begin{eqnarray}
{\cal L}_{eff} &=&i Z_b^{L} \overline{b_{L}}\gamma^{\mu}\left(
\partial_{\mu}-i\frac{s_0^2}{3}{\cal A}_{\mu}\right ) b_{L}
+ iZ_b^{R} \overline{b_{R}}\gamma^{\mu}\left(
\partial_{\mu}-i\frac{s_0^2}{3}{\cal A}_{\mu}\right ) b_{R}\nonumber \\
&&+\frac{1}{2}\left( Z_{v}^{L}-Z_{b}^{L}\frac{2s_0^2}{3}\right )
{{\overline{b_L}\gamma_{\mu} b_{L}{\cal Z}^{\mu}}\,}
-Z_b^{R}\frac{s_0^2}{3}{{\overline{b_R}\gamma_{\mu} b_{R}
{\cal Z}^{\mu}}\,} +{\dots} \,\, .
\label{eq38}
\end{eqnarray}
This effective Lagrangian summarizes all the loop corrections
in powers of $m_t$ in
the coefficient functions $Z_{b}^{L}$, $Z_{b}^{R}$,
and $Z_{v}^{L}$. Recall that up to now
all the fields in ${\cal L}_{eff}$ are bare fields.
To compare with the
low energy data we prefer to express ${\cal L}_{eff}$ in terms of
the renormalized fields.
In Eq.~(\ref{eq38}), the kinetic terms of
the $b_L$ and $b_R$ fields can be properly normalized
after redefining (renormalizing) the fields $b_L$ and $b_R$ by
${(Z_{b}^L)}^{\frac{-1}{2}} b_{L}$ and
${(Z_{b}^R)}^{\frac{-1}{2}} b_{R}$, respectively.
In terms of the renormalized fields $b_L$ and $b_R$,
${\cal L}_{eff}$ can be rewritten as
\begin{eqnarray}
{\cal L}_{eff} &=& \overline{b_{L}}i\gamma^{\mu}\left(
\partial_{\mu}-i\frac{s_0^2}{3}{\cal A}_{\mu}\right ) b_{L}
+ \overline{b_{R}}i\gamma^{\mu}\left(
\partial_{\mu}-i\frac{s_0^2}{3}{\cal A}_{\mu}\right ) b_{R}\nonumber \\
&&+\frac{1}{2}\left( \frac{Z_{v}^{L}}{Z_b^L} -
\frac{2s_0^2}{3}\right ){{\overline{b_L}\gamma_{\mu} b_{L}{\cal Z}^{\mu}}\,}
-\frac{s_0^2}{3}{{\overline{b_R}\gamma_{\mu} b_{R}
{\cal Z}^{\mu}}\,} +{\dots} \,\,\, .
\label{eq39}
\end{eqnarray}

Before considering the physical observables at low energy
let us first examine the bosonic sector.
Similar to our previous discussions,
loop corrections to the bosonic sector can be organized using
the effective Lagrangian
\begin{eqnarray}
{\cal L}^{B}_{eff} &=&-\frac{1}{4g_0^2} {{\cal W}_{\mu \nu}^a}
{{\cal W}^{\mu \nu}}^{a}
 -\frac{1}{4{g_0^\prime}^2} {\cal B}_{\mu \nu}
{\cal B}^{\mu \nu}\nonumber \\
&&+Z^{\phi}\frac{v_0^2}{4}{\cal W}^{+}_{\mu}{{\cal W}^{-}}^{\mu}+
Z^{\chi}\frac{v_0^2}{8}{\cal Z}_{\mu}{\cal Z}^{\mu}+{\dots }
\label{eq5}\,\, .
\end{eqnarray}
Note that in the above equation we have explicitly
used the subscript
$0$ to indicate bare quantities.
The bosonic Lagrangian in Eq.~(\ref{eq4})
and the identity in Eq.~(\ref{eq05}) imply that
the Yang-Mills terms (the first two terms in ${\cal L}^B$)
are not directly related to the SSB sector.
Hence, any radiative corrections to the field ${W}_{\mu \nu}^a$
must know about the weak coupling $g$, i.e., suppressed by $g$ in our
point of view. This also holds for operators, of dimension
four or higher, which include ${W}_{\mu \nu}^a$ in the chiral Lagrangian
where all these gauge invariant terms are suppressed by the
weak coupling $g$ ~\cite{app,fer}.
The same conclusion applies to $B_{\mu \nu}$.
Therefore we conclude that the fields ${\cal W}_{\mu}^{\pm},
{\cal Z}_{\mu}$, and ${\cal A}_{\mu}$ in
${\cal L}_{eff}$ and ${\cal L}^B_{eff}$ do not
get wavefunction corrections
(renormalization) in the limit of ignoring corrections
of the order $g$, namely the renormalized fields and the
bare fields are identical in this limit.

Expanding the mass terms in Eq.~(\ref{eq5}) we find
\begin{eqnarray}
Z^{\phi}\frac{v_0^2}{4}{\cal W}^{+}_{\mu}{{\cal W}^{-}}^{\mu}+
Z^{\chi}\frac{v_0^2}{8}{\cal Z}_{\mu}{\cal Z}^{\mu}
&=&Z^{\phi}\partial_{\mu}\phi^{+}\partial^{\mu}\phi^{-}
+\frac{1}{2}Z^{\chi}\partial_{\mu}\phi^{3}
\partial^{\mu}\phi^{3}\nonumber \\
&&+Z^{\phi}\frac{g_0^2v_0^2}{4}W_{\mu}^{+}{W^{-}}^{\mu}
+Z^{\chi}\frac{g_0^2v_0^2}{8 c_0^2}Z_{\mu}Z^{\mu}
+{\dots}\,\, .
\label{eq5prime}
\end{eqnarray}
It is clear that $Z^{\phi}$ denotes the self energy correction of
the charged Goldstone boson $\phi^{\pm}$, and $Z^{\chi}$ denotes
the self energy correction of the neutral Goldstone boson $\phi^3$.
Since $W^{\pm}_{\mu}$ and $Z_{\mu}$ do not get wavefunction correction in
powers of $m_t$, therefore the gauge boson masses are
\begin{eqnarray}
M_W^2 & = &Z^{\phi} \frac{g_0^2 v_0^2}{4} =Z^{\phi} {M^2_W}_{0} \,\, ,
\nonumber \\
M_Z^2 & = & Z^{\chi} \frac{g_0^2 v_0^2}{4c_0^2}=Z^{\chi} {M^2_Z}_{0} \,\, .
\label{wzmass}
\end{eqnarray}

In summary, all the loop corrections in powers of $m_t$ to low energy data
can be organized in the sum of ${\cal L}_{eff}$
[in Eq.~(\ref{eq39})] and ${\cal L}^B_{eff}$ [in Eq.~(\ref{eq5})].
Comparing them to the bare Lagrangian ${\cal L}_0$
in Eq.~(\ref{eq2}), we find that in the limit of taking $g \rightarrow 0$ the
chiral Lagrangian ${\cal L}_0$ behaves as a renormalizable theory
although in general a chiral Lagrangian is nonrenormalizable.
In other words, no higher dimensional operators (counterterms)
are needed to renormalize the theory in this limit.
The same feature was also found in another application of
a chiral Lagrangian with $1/N$ expansion \cite{largen}.

\subsection{Renormalization}
\indent

Now we are ready to consider the large $m_t$ corrections
to low energy data. We choose our renormalization scheme
to be the $\alpha$, $G_F$, and $M_Z$ scheme. With
\begin{equation}
g_{0}^2=\frac{4\pi \alpha_0}{s_0^2}\,\,
\end{equation}
and
\begin{equation}
s_0^2c_0^2=\frac{\pi \alpha_0}{\sqrt{2}{G_F}_{0}{M_Z^2}_{0}} \,\, ,
\end{equation}
or,
\begin{equation}
s_0^2=\frac{1}{2}\left[1-\left (1-\frac{4\pi\alpha_0}
{\sqrt{2}{G_F}_{0}{M_Z^2}_{0}}\right )^{1/2}\right]\,\, .
\end{equation}
Define the counterterms as
\begin{eqnarray}
\alpha &=& \alpha_0 + \delta \alpha \nonumber \,\, ,\\
G_F &=& {G_F}_0 + \delta G_F  \nonumber \,\, ,\\
M_Z^2 &=& {M_Z^2}_0 + \delta M_Z^2 \,\, ,
\end{eqnarray}
and
\begin{eqnarray}
s^2 &=& s_0^2 + \delta s^2 = s_0^2-\delta c^2
\nonumber \,\, ,\\
c^2 &=& c_0^2 + \delta c^2 \,\, ,
\end{eqnarray}
then
\begin{equation}
s^2c^2+(c^2-s^2) \,  \delta c^2=
\frac{\pi \alpha}{\sqrt{2}{G_F}{M_Z^2}} \left(
1 - \frac{\delta \alpha}{\alpha} + \frac{\delta G_F}{G_F}
 +\frac{\delta M_Z^2}{M_Z^2} \right)
\label{sinw} \,\, .
\end{equation}
As shown in the above equation,
even after the counterterms $\delta \alpha$, $\delta G_F$,
and $\delta M_Z^2$
are fixed by data [e.g., the electron (g-2), muon lifetime, and the
mass of the $Z$ boson], we still have the freedom to choose
$\delta c^2$ by using a different definition of the
renormalized quantity $s^2c^2$.
In our case we would  choose the definition of the
renormalized $s^2$ such that there will be no large top quark mass
dependence (in powers of $m_t$) in the counterterm $\delta c^2$.
We shall show later that for this purpose our renormalized $s^2$
satisfies \footnote{
If we define
$ s'^2c'^2 = \frac{\pi \alpha}{\sqrt{2}{G_F}{M_Z^2}} $, then
$s^2=s'^2 ( 1 + \Delta k')$ with
$ \Delta k'= -\frac{c'^2 \delta \rho }{c'^2 - s'^2} $,
and the counterterm of $s'^2$ will contain contributions in
powers of $m_t$.}
\begin{equation}
s^2c^2 = \frac{\pi \alpha}{\sqrt{2}{G_F}{M_Z^2} \rho} \,\, ,
\end{equation}
where $\rho$ is defined from the partial width of $Z$ into
lepton pairs, cf. Eq.~(\ref{widthmu}).
With this choice of $s^2$ and the definition of the renormalized
weak coupling
\begin{equation}
g^2=\frac{4 \pi \alpha}{s^2} \,\, ,
\end{equation}
one can easily show that the counterterm $\delta g^2$
($=g^2-g_0^2$) does not contain large $m_t$ dependence.
(Obviously, $\delta \alpha$ will not have contributions
purely in powers of $m_t$.)
Namely, in this renormalization scheme, $\alpha$, $g$,
and $s^2$ do not get renormalized after ignoring all the contributions
of the order $g$.
Hence, all the bare couplings $g_0$, $g_0'$, and $s_0^2$ in
the effective Lagrangians ${\cal L}_{eff}$ and ${\cal L}^B_{eff}$
do not get corrected when considering the contributions which
do not vanish in the limit of $g \rightarrow 0$.
The only non-vanishing counterterm needs to be considered
in Eq.~(\ref{eq5}) is $\delta v^2$ ($=v^2-v_0^2$).
{}From Eq.~(\ref{wzmass}) and $M_W=gv/2$, we find
\begin{equation}
Z^\phi v_0^2 = v^2 \,\, ,
\end{equation}
because neither $g$ nor ${\cal W}^\pm$
(or $W^\pm$) gets renormalized.
Thus,
\begin{equation}
{G_F}_0=\frac{1}{\sqrt{2} v^2_0}=
Z^{\phi} \frac{1}{\sqrt{2} v^2} = Z^{\phi} G_F \,\, .
\end{equation}
Consequently,
\begin{equation}
\frac{g_0^2}{c_0^2}=\frac{8{G_F}_{0}{M_Z^2}_{0}}{\sqrt{2}}
=\frac{8{G_F}{M_Z^2}}{\sqrt{2}}\frac{Z^\phi}{Z^{\chi}}\,\, ,
\end{equation}
and the effective $Z$-$b$-$b$ coupling is
\begin{equation}
-\frac{g_0}{2c_0}\gamma^{\mu}\left[ \left(
\frac{Z_{v}^L}{Z_{b}^{L}} -\frac{2s_0^2}{3}
\right)P_L -\frac{2s_0^2}{3}P_R\right]=\\
-{\sqrt{\frac{G_FM_Z^2}{2\sqrt{2}}\frac{Z^{\phi}}{Z^{\chi}}}}
\gamma^{\mu}\left[ \left(\frac{Z_{v}^L}{Z_{b}^{L}}
-\frac{4s^2}{3}\right) -\frac{Z_{v}^L}{Z_{b}^{L}}\gamma_5\right] \,\, ,
\label{effzbb}
\end{equation}
where $P_{L,R}=(1\mp\gamma_5)/{2}$.

\subsection{Low Energy Observables}
\indent

In general, all the radiative corrections to low energy data can be
categorized in a model independent way into four parameters:
$S$, $T$, $U$ \cite{peskin}, and $R_b$ \cite{joserb};
or equivalently,
$\epsilon_1$, $\epsilon_2$, $\epsilon_3$, and $\epsilon_b$ \cite{bar}.
The relations between these two sets of parameters are, to the order
of interest,
\begin{eqnarray}
S &=& \frac{4 s^2}{\alpha(M_Z^2)} \epsilon_3 \,\, , \nonumber \\
T &=& \frac{1}{\alpha(M_Z^2)} \epsilon_1 \,\, , \nonumber \\
U &=& -\frac{4 s^2}{\alpha(M_Z^2)} \epsilon_2 \,\, ,
\end{eqnarray}
and both $R_b$ ($=\Gamma_b / \Gamma_h$)
 and $\epsilon_b$ measure the
effects of new physics in the partial decay width ($\Gamma_b$) of
$Z \rightarrow b \overline b$. ($\Gamma_h$ is the hadronic width of $Z$.)

  These parameters can be
 derived from four basic measured \mbox{observables}, such as
$\Gamma_{\mu}$\,(the partial decay width of $Z$ into a
$\mu$ pair),
$A_{FB}^{\mu}$\,(the forward-backward asymmetry at the $Z$ peak for
the $\mu$ lepton), $ M_{W} / M_{Z} $ (the ratio of
$W^\pm$ and $Z$ masses),
and $\Gamma_{b}$\,(the partial decay width
of $Z$ into a $b\overline{b}$ pair).
The expressions of these observables in terms of
$\epsilon$'s can be found in Ref.~\cite{bar}.

In this paper we only give the relevant terms in $\epsilon$~'s
that might contain the leading effects in powers of $m_t$
from new physics.
Denote the vacuum polarization for the $W^1$, $W^2$,
$W^3$, and $B$ gauge bosons as
\begin{equation}
{{\Pi}^{ij}}_{\mu \nu}(q) = -ig_{\mu \nu}
\left [A^{ij}(0) + q^{2}F^{ij}(q^2)
\right ] + q_{\mu}q_{\nu}\,\rm{terms}\, ,
\end{equation}
where $i,j=1,2,3,0$ for $W^1, W^2, W^3$ and $B$,  respectively. Then,
\begin{equation}
\epsilon_1 = e_1-e_5 \,,
\end{equation}
\begin{equation}
\epsilon_2 = e_2 - c^{2}e_5\, ,
\end{equation}
\begin{equation}
\epsilon_3 = e_3 - c^{2}e_5 \, ,
\end{equation}
\begin{equation}
\epsilon_b = e_b \, ,
\end{equation}
where
\begin{equation}
e_1 = \frac{A^{33}(0) - A^{11}(0)}{{M^2_W}}\, ,
\end{equation}
\begin{equation}
e_2 = F^{11}({M^2_W}) - F^{33}({M^2_Z})\, ,
\end{equation}
\begin{equation}
e_3 = \frac{c}{s}F^{30}({M^2_Z})\, ,
\end{equation}
\begin{equation}
e_5 = {M^2_Z}\frac{dF^{ZZ}}{dq^2}({M^2_Z})\, ,
\end{equation}
and $e_b$ is defined through the Glashow-Iliopoulos-Maiani-
(GIM-) violating
$Z\rightarrow b\overline{b}$ vertex
\begin{equation}
{V_{\mu}}^{{\rm {GIM}}}\left ( Z \rightarrow b\bar{b}\right ) =
-\frac{g}{2c}e_b \gamma_{\mu}\frac{1-\gamma_5}{2}\, .
\end{equation}

Both $\epsilon_1$ and $\epsilon_b$ gain corrections in powers of
$m_t$ \cite{ehab}, and are sensitive to
new physics coming through the top quark.
On the contrary, $\epsilon_2$\, and $\epsilon_3$ do not play any
significant role in our analysis because their dependence on the top mass is
only logarithmic.
Hence,
\begin{eqnarray}
\epsilon_1 &=& \delta \rho +
{\rm{corrections \,\, of \,\, the \,\, order }}\,\, g\, ,
\nonumber \\
\epsilon_b &=& \tau +
{\rm{corrections \,\, of \,\, the \,\, order }}\,\, g\, ,
\nonumber \\
\epsilon_2 &=&
{\rm{corrections \,\, of \,\, the \,\, order }}\,\, g\, ,
\nonumber \\
\epsilon_3 &=&
{\rm{corrections \,\, of \,\, the \,\, order }}\,\, g\, ,
\end{eqnarray}
where $\delta \rho = \rho -1$.
The parameters $\rho$ and $\tau$ are defined by
\begin{eqnarray}
\Gamma_{\mu} & \equiv & \Gamma(Z\rightarrow \mu^{+}\mu^{-})=\rho
\frac{G_F M_Z^3}{6\pi\sqrt{2}}
\left ({g_{\mu}^2}_V+ {g_{\mu}^2}_A \right )
\label{widthmu} \,\, , \nonumber \\
\Gamma_b & \equiv & \Gamma(Z\rightarrow \overline{b}b)=\rho
\frac{G_F M_Z^3}{2\pi\sqrt{2}}
\left ({g_{b}^2}_V+ {g_{b}^2}_A \right )\,\, ,
\label{gammamub}
\end{eqnarray}
where
\begin{eqnarray}
{g_{\mu}}_V & = &- \frac{1}{2} \left( 1-4s^2 \right)
,\,\,\, {g_{\mu}}_A=-\frac{1}{2} \,\, ,  \nonumber \\
{g_{b}}_V & = & -\frac{1}{2} \left(1-\frac{4}{3}s^2+\tau \right)
  ,\,\,\,  {g_{b}}_{A}=-\frac{1}{2} \left( 1+\tau \right) \,\, .
\end{eqnarray}
Hence, comparing to Eq.~(\ref{effzbb}) we conclude
\begin{eqnarray}
\delta\rho & = &\frac{Z^{\phi}}{Z^{\chi}}-1\,\, , \nonumber \\
\tau & =&\frac{Z_{v}^{L}}{Z_{b}^{L}}-1
\,\, .
\label{rhotauall}
\end{eqnarray}

\subsection{One Loop Corrections in the SM}
\indent

The SM, being a linearly realized ${\rm{SU(2)}}_L \times {\rm{U(1)}}_Y$
gauge theory, can be formulated as a chiral Lagrangian
after nonlinearly transforming the fields \cite{ehab}.
Applying the previous formalism, we  calculate the one-loop corrections
of order $m_t^2$ to $\rho$ and $\tau$ for the SM by
taking $\kappa_{L}^{\rm {NC}}=\kappa_{R}^{\rm {NC}}=
\kappa_{L}^{\rm {CC}}=\kappa_{R}^{\rm {CC}}=0$ in Eq.~(\ref{eq2}).
These loop corrections can be summarized by the
coefficient functions
$Z^{\chi}$, $Z^{\phi}$, $Z^L_b$, and $Z^L_v$ which are calculated
from the Feynman diagrams shown in Figs.~1(a), 1(b), 1(c),
and the sum of 1(d) and 1(e), respectively.
We find
\begin{eqnarray}
Z^{\chi}&=&1+\frac{6m_t^2}{16\pi^2 v^2}\left (
\Delta-\ln m_t^2\right) \nonumber\,\, , \\
Z^{\phi}&=&1+\frac{6m_t^2}{16\pi^2 v^2}\left (\Delta+\frac{1}{2}-
\ln m_t^2\right)\nonumber  \,\, ,\\
Z_{b}^L&=&1+\frac{3m_t^2}{16\pi^2 v^2}\left (
-\Delta+\ln m_t^2 -\frac{5}{6}\right)\,\, ,\nonumber\\
Z_{v}^L&=&1+\frac{3m_t^2}{16\pi^2 v^2}\left (-\Delta+\ln m_t^2 -\frac{3}{2}
\right ) \,\, .
\end{eqnarray}
We note that Fig. 1(e) arises from the nonlinear realization of the
gauge symmetry in the chiral Lagrangian approach.
Substituting the above results into Eq.~(\ref{rhotauall}), we obtain
\begin{eqnarray}
\delta \rho &=& \frac{3G_F m_t^2}{8\sqrt{2}\pi^2}\,\, , \nonumber \\
\tau &=& -\frac{G_F m_t^2}{4\sqrt{2}\pi^2}
\,\, ,
\end{eqnarray}
which are the established results~\cite{bar2}.

\section{Constraining the top quark couplings to the EW gauge bosons}
\indent

In Ref.~\cite{ehab} we calculated the one-loop corrections
(of order $m_t^2\ln \Lambda^2)$ to $\rho$ and $\tau$
due to the nonstandard couplings of the top quark to the EW
gauge bosons. The set of Feynman diagrams we considered contained
external massive gauge bosons lines.
In this paper we show how to reproduce those results
by considering a set of Feynman diagrams which contains only
the pure Goldstone bosons, the top quark, and the bottom quark lines,
as described in Sec. 2.

Non-renormalizability of the effective Lagrangian presents
a major problem on how to find a scheme to
handle both the divergent and the finite pieces in
loop calculations \cite{burg}. Such a problem arises because
the underlying theory is not yet known, so it is not possible to
apply the exact matching conditions
to find the correct scheme to be used in the effective Lagrangian
\cite{geor}.
One approach is to associate the divergent piece in
loop calculations with a physical
cutoff $\Lambda$, the upper
scale at which the effective Lagrangian is
valid \cite{pecc}. In the chiral Lagrangian approach this cutoff
$\Lambda$ is taken to be
$4\pi v \sim 3$\,TeV \cite{geor}.\footnote{
This scale, $4\pi v \sim 3$\,TeV, is only meant to indicate
the typical cutoff scale. It is equally probable to have, say,
$\Lambda=1$\,TeV.} For the finite piece no
completely satisfactory approach is available \cite{burg}.

To perform loop calculations using the chiral Lagrangian,
one should arrange
the corrections in powers of $1/4\pi v$ and include all
the Feynman diagrams up to the desired order.
Fig.~1 contains all the Feynman diagrams needed for
our study.
We calculate the leading  contribution to $\rho$ and
$\tau$ due to the new interaction terms in the chiral Lagrangian
using the dimensional regularization scheme and
taking the bottom quark mass to be zero.
At the end of the calculation, we
replace the divergent piece $1/\epsilon$ by
$\ln(\Lambda^2/{m_t^2})$ for $\epsilon = (4-n)/2$, where $n$ is the
space-time dimension. Effectively,
we have assumed that the underlying full theory is renormalizable.
The cutoff scale $\Lambda$ serves as the infrared cutoff of the
operators in the effective Lagrangian. Due to the renormalizability
of the full theory, from renormalization group analysis, we conclude
that the same cutoff $\Lambda$ should also serve as the ultraviolet
cutoff of the effective Lagrangian in calculating Wilson coefficients.
Hence, in the dimensional regularization scheme,
$1/\epsilon$ is replaced by $\ln(\Lambda^2/{\mu^2})$.
Furthermore, the renormalization scale $\mu$ is set to be $m_t$,
the heaviest mass scale in the effective Lagrangian of interest.
Since we are mainly interested
in new physics associated with the top quark couplings to gauge bosons,
we shall restrict ourselves
to the {\it leading} contribution enhanced by the top quark mass, i.e.,
of the order of $\left (m_t^{2}\ln {\Lambda}^{2}\right)$.

Inserting these nonstandard couplings in loop diagrams and keeping only
the linear terms in $\kappa$'s,
we find
\begin{eqnarray}
Z^{\chi}&=&1+\frac{6m_t^2}{16\pi^2 v^2}
\left(2\kappa_{L}^{\rm {NC}}-2\kappa_{R}^{\rm {NC}}\right)
\ln \frac{\Lambda^2}{m_t^2} \nonumber\,\, , \\
Z^{\phi}&=&1+\frac{12m_t^2}{16\pi^2 v^2}\kappa_{L}^{\rm {CC}}
\ln \frac{\Lambda^2}{m_t^2} \nonumber\,\, ,\\
Z_{b}^L&=&1-\frac{6m_t^2}{16\pi^2 v^2}\kappa_{L}^{\rm {CC}}\ln
\frac{\Lambda^2}{m_t^2}\nonumber \,\, ,\\
Z_{v}^L&=&1-\frac{m_t^2}{16\pi^2 v^2}
\left(6\kappa_{L}^{\rm {CC}}-4\kappa_{L}^{\rm {NC}}+
\kappa_{R}^{\rm {NC}}\right)
\ln \frac{\Lambda^2}{m_t^2} \,\, .
\end{eqnarray}
Thus the nonstandard contributions to $\rho$ and $\tau$ are
\begin{eqnarray}
\delta \rho &=& \frac{3G_F m_t^2}{2\sqrt{2}\pi^2}\left (
\kappa_{L}^{\rm {CC}}-\kappa_{L}^{\rm {NC}}+\kappa_{R}^{\rm {NC}}\right)
\ln \frac{\Lambda^2}{m_t^2} \,\, ,  \nonumber \\
\tau &=& \frac{G_F m_t^2}{2\sqrt{2}\pi^2}\left (
-\frac{1}{4}\kappa_{R}^{\rm {NC}}+\kappa_{L}^{\rm {NC}}\right)
\ln \frac{\Lambda^2}{m_t^2}\,\, ,
\label{rhotau}
\end{eqnarray}
which agree with our previous results obtained in Ref.~\cite{ehab}.

In Ref.~\cite{zhang} a similar calculation for $\tau$ was performed and the
author claimed to get a different result from ours.
However, the author included only the vertex corrections
to calculate the physical quantity $\tau$, which according to our
systematic discussion in the previous section
is not complete because the wavefunction corrections to
the $b$ quark must be included.

Based upon the new LEP measurements \cite{lep94},
a global analysis indicates a SM top quark mass to be \cite{alt}
\begin{equation}
m_t=165 \pm 12\,\, {\rm{GeV\,\,\, for}}\,\,\, m_H=300 \,\,{\rm{GeV}}\,\,.
\end{equation}
If the SLC measurement is included with LEP measurements, then
\begin{equation}
m_t=174 \pm 11\,\, {\rm{GeV\,\,\, for}}\,\,\, m_H=300 \,\,{\rm{GeV}}\,\,.
\end{equation}

Using the new LEP and SLC results we shall update the constraints on the
nonstandard couplings of the top quark to the EW gauge bosons.
This can be done
by comparing the new experimental values for $\delta \rho$ and $\tau$ with
that predicted by the SM and the nonstandard contributions combined.
In the limit of ignoring the contributions of the order $g$,
the observables $\Gamma_\mu$, $A_{FB}^\mu$, $M_W / M_Z$, and $\Gamma_b$
can all be expressed in terms of the two quantities
$\delta \rho$ and $\tau$.
In addition to Eq.~(\ref{gammamub}), we find
\begin{equation}
A_{FB}^{\mu}= \frac{ 3 {g_{\mu}^2}_V {g_{\mu}^2}_A }
{ \left( {g_\mu^2}_V + {g_\mu^2}_A \right)^2 } \,\,
\end{equation}
and\footnote{
In terms of the quantity $\Delta r_w$ defined in Ref.~\cite{bar},
$\frac{M_W^2}{M_Z^2} \left( 1 - \frac{M_W^2}{M_Z^2} \right)
= \frac{ \pi \alpha(M_Z^2) }
{ \sqrt{2} G_F M_Z^2 ( 1-\Delta r_w) } $. For corrections in powers
of $m_t$, $ s^2 \Delta r_w = - c^2 \delta \rho $.
}
\begin{equation}
\frac{M_W^2}{M_Z^2} = \rho \, c^2 \,\, .
\end{equation}

Using the minimum set of observables
($\Gamma_\mu$, $A_{FB}^\mu$, $M_W / M_Z$, and $\Gamma_b$),
we constrain the allowed space of $\kappa$'s in
a model independent way, i.e., without specifying the explicit dynamics
for generating these nonstandard effects.
One can also enlarge the set of observables used in the analysis by
including all the LEP data and
the SLC measurement of the left-right cross section asymmetry $A_{LR}$
in $Z$ production with a longitudinally polarized electron beam,
 where
\cite{bar}
\begin{equation}
A_{LR} = \frac{2x}{1+x^2} \, ,
\end{equation}
for
\begin{equation}
x=  \frac{ {g_e}_V }{ {g_e}_A } = 1-4 s^2  \,\, .
\end{equation}

Following the same analyses carried out in Ref.~\cite{ehab}, we
include both the SM and the nonstandard contributions to low energy data.
The SM contributions to $\delta \rho$ and $\tau$ were given in
Ref.~\cite{bar} for various top quark and
Higgs boson masses. Our conclusions are however
not sensitive to the Higgs boson mass \cite{ehab}.

Choosing $m_t=175$\,GeV and $m_H=100$\,GeV, we span the parameter space
defined by $-1.0\leq \kappa_{L}^{\rm {NC}}\leq 1.0$,
$-1.0\leq \kappa_{R}^{\rm {NC}} \leq 1.0$, and
$-1.0\leq\kappa_{L}^{\rm {CC}}\leq 1.0$,
and compare with the values\footnote{
$\epsilon_1=\delta \rho$, $\epsilon_b=\tau$,
$\epsilon_2=(-9.2 \pm 5.1)\times 10^{-3}$, and
$\epsilon_3=(3.8 \pm 1.9)\times 10^{-3}$.}
\begin{equation}
\delta \rho =( 3.5\pm 1.8 )\times 10^{-3}\,\, ,
\end{equation}
\begin{equation}
\tau=(0.9\pm 4.2)\times 10^{-3}
\end{equation}
from a global fit \cite{alt} using all the new LEP and SLC data.
For reference, we list here some of the relevant data,
taken from \cite{alt},
\begin{eqnarray}
{\alpha}^{-1}(M_Z^2) &=&   128.87 \pm 0.12  \nonumber\, ,\\
G_F          & =  & 1.16637(2) \times {10}^{-5} \,\,\,\,
 {\rm{GeV}}^{-2} \nonumber\, ,\\
M_Z          & =  &  91.1899 \pm 0.0044  \,\,\, \rm{GeV} \nonumber \, ,\\
M_W/M_Z      & =  &  0.8814 \pm 0.0021   \nonumber \, ,\\
\Gamma_{\ell}& =  &  83.98 \pm 0.18  \,\,\, \rm{MeV} \nonumber \, ,\\
\Gamma_b     & =  &  385.9 \pm 3.4  \,\,\,\rm{MeV}   \nonumber \, ,\\
A^{\ell}_{FB}       & =  &  0.0170 \pm 0.0016 \nonumber \, ,\\
A^{b}_{FB}       & =  &  0.0970 \pm 0.0045  \nonumber \, ,\\
A_{LR} \, ({\rm SLC}) \,       & =  &  0.1668 \pm 0.079 \nonumber \, .
\end{eqnarray}

We find that within $2\sigma$ the allowed
region of these three parameters exhibits the same features as that
obtained using the old set of data (see Ref.~\cite{ehab}).
These features can be deduced from the two-dimensional projections of
the allowed parameter space, as shown in Figs.~2, 3, and 4.
They are briefly summarized as follows:
\begin{itemize}
\item[(1)]
As a function of the top quark mass, the allowed parameter space shrinks
as the top quark mass increases.
\item[(2)]
Data do not exclude possible
new physics coming through the top quark couplings
to the EW gauge bosons.
As shown in Fig.~2,  $\kappa_{L}^{\rm {CC}}$ and
$\kappa_{R}^{\rm {NC}}$ are not yet constrained by the current data.
Furthermore, no conclusion can be drawn about
$\kappa_{R}^{\rm {CC}}$ because $\kappa_{R}^{\rm {CC}}$
does not contribute to the LEP or the SLC observables in
the limit of taking $m_b = 0$.
\item[(3)]
$\kappa_{L}^{\rm {NC}}$ is almost constrained. New physics prefers
positive $\kappa_{L}^{\rm {NC}}$, see Figs.~3 and 4. For example,
$\kappa_{L}^{\rm {NC}}$ is
constrained within ($-0.3$ to 0.5) for a 175 GeV top quark.
\item[(4)]
New physics prefers $\kappa_{L}^{\rm {CC}}\approx -\kappa_{R}^{\rm {NC}}$.
This is clearly shown in Fig.~2.
\end{itemize}

As compared with the old set of data from LEP and SLC,  new data
tighten the allowed region of the nonstandard parameters $\kappa$'s by
no more than a factor of two. This difference is due
to the slightly smaller
errors on the new measurements as compared with the old ones.
The largest impact of these new data on our results
comes from the more precise measurement of
$\Gamma_b$  which turns out to be about
$2\sigma$ higher than the SM prediction and implies a lighter
top quark. For a much heavier top quark, new physics must come in because
all the $\kappa$'s cannot simultaneously vanish.
If the large discrepancy between LEP and SLC data persists, then
our model of having nonstandard top quark couplings to the
gauge bosons is one of the candidates that can accommodate
such a difference.

If we restrict ourselves to the minimum set of observables,
which give \cite{alt}
\begin{equation}
\delta \rho =(4.8\pm 2.2)\times 10^{-3}\,\, ,
\end{equation}
\begin{equation}
\tau=(5.0\pm4.8)\times 10^{-3}\,\, ,
\end{equation}
we reach almost the same conclusion. The
main difference is that $\kappa_{L}^{\rm {NC}}$ shifts slightly to the right,
due to the fact that the central value of
$\tau$ in this case is larger than its global fit value.

In Ref.~\cite{ehab} we discussed an effective model incorporated
with an additional approximate custodial symmetry
(responsible for $\rho=1$ at the tree-level), and
concluded that $\kappa_{L}^{\rm {NC}}=2\kappa_{L}^{\rm {CC}}$ as
long as the tree-level vertex
{\mbox{$b$-$b$-$Z$}} is not modified.
{}From Eq.~(\ref{rhotau}), we find for this model
\begin{equation}
\delta \rho=\frac{3G_F m_t^2}{2\sqrt{2}\pi^2}\left (
-\frac{1}{2}\kappa_{L}^{\rm {NC}}
+\kappa_{R}^{\rm {NC}}\right)\ln \frac{\Lambda^2}{m_t^2}\,\,
\end{equation}
and
\begin{equation}
\tau =\frac{G_F m_t^2}{2\sqrt{2}\pi^2}\left (-\frac{1}{4}
\kappa_{R}^{\rm {NC}}+\kappa_{L}^{\rm {NC}}\right)
\ln \frac{\Lambda^2}{m_t^2}\,\, .
\end{equation}
Using this effective model, we span the plane
defined by $\kappa_{L}^{\rm {NC}}$ and $\kappa_{R}^{\rm {NC}}$ for top quark
mass 150\,GeV and 175\,GeV, respectively.
Figs.~5 and 6 show the allowed range for those parameters within
$2\sigma$. As a general feature one observes that the allowed range forms
a narrow area aligned close to the line $\kappa_{L}^{\rm {NC}}=2
\kappa_{R}^{\rm {NC}}$. For
$m_t=150$\,GeV (175 GeV) we see that
$-0.05\leq \kappa_{L}^{\rm {NC}}\leq 0.3$
($0.0\leq \kappa_{L}^{\rm {NC}}\leq 0.25$). As the top quark mass increases
this range shrinks and moves downward to the right, away from the origin
$(\kappa_{L}^{\rm {NC}},\kappa_{R}^{\rm {NC}})=(0,0)$, although
positive $\kappa$'s remain preferred.
 The reason for this behavior is simply due to
the fact that as $m_t$ increases, the SM value for $\rho$ ($\tau$)
increases in the positive (negative) direction.
To summarize this behaviour, we show, respectively, in Figs.~7 and 8
the allowed ranges for $\kappa_{L}^{\rm {CC}}$ and
$(\kappa_{L}^{\rm {NC}} - 2 \kappa_{R}^{\rm {NC}})$ as a function of $m_t$.
An interesting point to mention is that in the global
fit analysis the SM ceases to be
a solution for $m_t \geq 200$\,GeV. However, with new physics effects,
e.g., $\kappa_{L}^{\rm {CC}}\neq 0$, $m_t$ can be as large as 300 GeV.

In this analysis we concentrated on  physics at the $Z$
resonance, i.e., at LEP and SLC.
Other low energy observables may as well be used to constrain
the nonstandard couplings of the top quark to the gauge bosons. In
Ref.~\cite{fuj} a constraint on the right-handed charged
current, $\kappa_{R}^{\rm {CC}}$, was set using the CLEO measurement of
$b\rightarrow s\gamma$. The authors
concluded that $\kappa_{R}^{\rm {CC}}$ is well constrained to within
a few percent
from its SM value ($\kappa_{R}^{\rm {CC}}=0$).
This provides a complementary information to our result
because LEP and SLC data are
not sensitive to $\kappa_{R}^{\rm {CC}}$ as compared to
$\kappa_{L}^{\rm {CC}}$, $\kappa_{L}^{\rm {NC}}$, and $\kappa_{R}^{\rm {NC}}$.

\section{Conclusions}
\indent

Because the top quark is heavy (close to the symmetry-breaking scale)
it will be more sensitive than the other light fermions
to new physics from the SSB sector.
Concentrating on effects, to low energy data, directly
related to the SSB sector, we took the chiral Lagrangian
approach to examine whether
the nonstandard couplings, $\kappa$'s, of the top quark to the
gauge bosons ($W^\pm$ and $Z$)
were already strongly constrained by
the old (1993) data from LEP and SLC ~\cite{ehab}.
Surprisingly, we found that to the order of
$\left ( m_t^2 \ln \Lambda^2\right )$
only the left-handed neutral current, $\kappa_{L}^{\rm {NC}}$, was
somewhat constrained
by the precision low energy data, although data
did impose some correlations among
$\kappa_{L}^{\rm {NC}}$, $\kappa_{R}^{\rm {NC}}$,
and $\kappa_{L}^{\rm {CC}}$. Since $\kappa_{R}^{\rm {CC}}$
does not contribute to the LEP or the SLC
observables in the limit $m_b = 0$,
$\kappa_{R}^{\rm {CC}}$ cannot be constrained by these data. However,
it was shown  in Ref.~\cite{fuj} that
$\kappa_{R}^{\rm {CC}}$ was already
constrained by the complementary process
$b \rightarrow s \gamma$.

In Ref.~\cite{ehab} we obtained our results by considering a set
of Feynman diagrams, derived form the nonlinear chiral Lagrangian,
whose external lines were the massive gauge boson lines.
The leading corrections (in power of $m_t$) to the low energy
observables were found not to vanish in the limit of
vanishing $g$ (the weak coupling)
because they originate from strong
couplings to the SSB sector, e.g., through large Yukawa coupling $g_t$.
Therefore, our previous results
should in principle be reproduced by considering an
effective Lagrangian which involves only the
scalar (the unphysical Goldstone bosons
and probably the Higgs boson) and the top-bottom fermionic sectors.
This was shown in  Sec. 2.
We discussed how to relate the two corresponding
sets of Green's functions for the low energy observables of interest.
We showed that by considering a completely different set of
Green's functions (without involving any external gauge boson line)
from that  discussed in Ref.~\cite{ehab} we
obtained exactly the same results. Our result for $\tau$
is different
from that given in Ref.~\cite{zhang} where the wavefunction
correction to the bottom quark was not included.

In Sec. 3 we used the new (1994) LEP and SLC data to constrain the
nonstandard interactions of the top quark to the EW gauge bosons.
As compared with the old (1993) data from LEP and SLC, the new data
tighten the allowed region of the nonstandard parameters, $\kappa$'s, by
no more than a factor of two. This difference is mainly due to
the more precise measurement of
$\Gamma_b$  which turns out to be about
$2\sigma$ higher than the SM prediction and favors a lighter top
quark.
If the large discrepancy between LEP and SLC data persists, then
our model of having nonstandard top quark couplings to the
gauge bosons is one of the candidates that can accommodate
such a difference. Positive values for $\kappa$'s are preferred for
the special model discussed in Ref.~\cite{ehab}, where an approximate
custodial symmetry is assumed.

\section*{ Acknowledgments }

We thank X. Zhang for drawing our attention to his work in
Ref.~\cite{zhang}. We also thank W. Repko for a critical reading
of the manuscript.
This work was supported in part by NSF Grant No. PHY-9309902.


\newpage
\section*{Figure Captions}
Fig. 1.\\
The relevant Feynman diagrams, which contribute to $\rho$ and
$\tau$ to the order $O({m_t^2}\ln {\Lambda}^2)$.
\label{fey}
\vspace{0.3cm}

\noindent
Fig. 2.\\
Two-dimensional projection in the plane of $\kappa_{R}^{\rm {NC}}$
and $\kappa_{L}^{\rm {CC}}$, for
$m_t=175$\,GeV and $m_H=100$\,GeV.
\vspace{0.3cm}

\noindent
Fig. 3.\\
Two-dimensional projection in the plane of $\kappa_{L}^{\rm {NC}}$
and $\kappa_{R}^{\rm {NC}}$, for
$m_t=175$\,GeV and $m_H=100$\,GeV.
\vspace{0.3cm}

\noindent
Fig. 4.\\
Two-dimensional projection in the plane of $\kappa_{L}^{\rm {NC}}$
and $\kappa_{L}^{\rm {CC}}$, for
$m_t=175$\,GeV and $m_H=100$\,GeV.
\vspace{0.3cm}

\noindent
Fig. 5.\\
The allowed region of
$\kappa_{L}^{\rm {NC}}$ and $\kappa_{R}^{\rm {NC}}$
($\kappa_{L}^{\rm {NC}} =2\kappa_{L}^{\rm {CC}}$), for
$m_t=150$\,GeV and $m_H=100$\,GeV. \label{f5}
\vspace{0.3cm}

\noindent
Fig. 6.\\
The allowed region
 of $\kappa_{L}^{\rm {NC}}$ and $\kappa_{R}^{\rm {NC}}$
($\kappa_{L}^{\rm {NC}} =2\kappa_{L}^{\rm {CC}}$), for
$m_t=175$\,GeV and $m_H=100$\,GeV. \label{f6}
\vspace{0.3cm}

\noindent
Fig. 7.\\
The allowed range of $\kappa_{L}^{\rm {CC}}$ as a
function of the mass of the top quark. (Note that
$\kappa_{L}^{\rm {NC}}=2\kappa_{L}^{\rm {CC}}$.)
\label{f7}
\vspace{0.3cm}

\noindent
Fig. 8.\\
The allowed range of $\kappa_{L}^{\rm {NC}} - 2 \kappa_{R}^{\rm {NC}}$ as a
function of the mass of the top quark.
(Note that $\kappa_{L}^{\rm {NC}}=2\kappa_{L}^{\rm {CC}}$.)
\label{f8}

\end{document}